\documentclass[preprint,12pt]{elsarticle}
\usepackage{amssymb}

\journal{Physica A}

\begin{document}

\begin{frontmatter}

\title{Recorded punishment promotes cooperation in spatial prisoner's dilemma game}

\author{Qing Jin$^{a}$, Zhen Wang$^{a,\ast}$, Zhen Wang$^{b,\ast}$, Yi-Ling Wang$^{c,\ast}$}

\address{$^a$ School of Physics, Nankai University, Tianjin, 300071, China\\
$^b$ School of Innovation Experiment, Dalian University of Technology, Dalian, 116024, China\\
$^c$ School of Life Science, Shanxi Normal University, LinFen, Shanxi 041000, China\\
{E-mail:zhenwang@mail.nankai.edu.cn; wangz@dlut.edu.cn; ylwangbj@yahoo.com.cn}}

\begin{abstract}
Previous studies suggest that punishment is a useful way to promote cooperation in the well-mixed public goods game, whereas it still lacks specific evidence that punishment maintains cooperation in spatial prisoner's dilemma game as well. To address this issue, we introduce a mechanism of recorded punishment, involved with memory and punishment, into spatial prisoner's dilemma game. We find that increasing punishment rate or memory length promotes the evolution of cooperation monotonously. Interestingly, compared with traditional version, recorded punishment will facilitate cooperation better through a recovery effect. Moreover, through examining the process of evolution, we provide an interpretation to this promotion phenomenon, namely, the recovery effect can be warranted by an evolution resonance of standard deviation of fitness coefficient. Finally, we confirm our results by studying the impact of uncertainty within strategy adoptions. We hope that our work may sharpen the understanding of the cooperative behavior in the society.\\
\end{abstract}

\begin{keyword}
Prisoner's Dilemma Game \sep Cooperation \sep
Punishment

\end{keyword}

\end{frontmatter}

\section{Introduction}

Cooperative behavior is abundant in the real world, ranging from microorganism groups to complex human societies \cite{1,2}. To understand the emergence and persistence of cooperation, it has attracted great interest in biology, physics, economics, as well as sociology \cite{3,4}. Evolutionary game theory has proved to be one of the most fruitful approaches to investigate this problem by studying evolutionary models based on the so-called social dilemmas \cite{5,6}. Well-known examples of these dilemmas include public goods game for group of interaction individuals \cite{7,8,9,10}, snowdrift game \cite{11,12} and prisoner's dilemma game \cite{13,14,15,16,17} as paradigms for pairwise interactions. Among these evolutionary games, the prisoner's dilemma game, in particular, has acquired prominent achievements in theoretical and experimental studies \cite{18,19,20}. In a typical prisoner's dilemma game, two players have a choice between cooperation and defection. They will receive the reward $R$ if both players cooperate, and the punishment $P$ if both choose defection. However, if one chooses defection against a cooperator, it will attain the temptation $T$ while the co-player obtain the sucker's payoff $S$. The ranking of four payoffs satisfies $T>R>P>S$, from which it is clear that selfish players are forced to choose defection which is best for individual, irrespective of the co-player's option, hence the dilemma occurs. In order to overcome this dilemma, specific mechanisms supporting cooperation are needed (see Ref. \cite{21} for a recent review).

Over the past decades, a number of mechanisms have been proposed which are able to support the emergence of cooperation \cite{22,23,24}, while the introduction of spatial structure is one of them, which is also refereed as network reciprocity \cite{25,26}. This successful research was firstly suggested by Nowak and May in their seminal paper \cite{20}. In the spatial game, players were situated on the vertices of a graph. Each player did not interact with every other, but with its neighbors that were marked by the direct edges. The payoff of each player was acquired by playing the game with its neighbors. Then the evolution of individual was determined through adopting the strategy of its neighbor, provided its fitness was higher. Interestingly, it proved that network reciprocity played a significant role in sustaining cooperation, and cooperators could survive through forming compact clusters. Stimulated by this pioneering work, a great many investigations based on spatial structure have been extended to date \cite{27,28,29,30} (for a survey see \cite{31}). Most notably, paradigmatic examples include heterogeneous activity of players \cite{32,33}, reward mechanism \cite{34}, influence of noise level \cite{27,35}, preferential selection of neighbors \cite{36}, effect of expected payoffs \cite{37}, mobility of players \cite{38,39,40}, differences in time scales \cite{41}, to name but a few. While among these mechanisms, what triggers our research interest most is the reward mechanism. In a recent work of Szolnoki and Perc \cite{42}, it shows that moderate rewards may promote cooperation better than high rewards. Meanwhile, we note that the promise of punishment and reward in promoting the evolution of cooperation is debatable, and the effect of punishment was mainly investigated in public goods game \cite{43,44,45}, but seldom studied in prisoner's dilemma game. Motivated by these facts, it is meaningful to consider a simple punishment mechanism in the prisoner's dilemma game, where players will be appropriately punished if they fail to pass their strategies to their offspring, since similar phenomena about punishment are ubiquitous in modern human and economy societies.

Furthermore, it is found that the coevolution of game theory is becoming a mushrooming avenue to explore the evolution of cooperation \cite{46,47,48,49}. Via coevolution, not only the evolution of strategies over time could be reflected, but also the adaptive development of network topologies or evolution rule (for a further view see Refs. \cite{50,51}). In recent investigations \cite{49}, they showed that cooperation could be promoted within a large scale when the coevolution of strategy updating and network topology was taken into account. In addition, we also note that when the teaching activity of players evolved with time, the maintenance of cooperation could be greatly promoted as well \cite{52}. Inspired by these innovations and the above idea of punishment, it is instructive to ask if we introduce memory length into punishment mechanism, namely, the punishment of each player changes over time, is it beneficial for the evolution of cooperation or not?

In this work, we study the spatial prisoner's dilemma game with the introduction of recorded punishment which is involved with memory and punishment. Before the game, each player was uniformly distributed the same fitness coefficient, which makes the game return to the traditional version. However, with the evolvement of the game, individual fitness coefficient will also change according to its current state and memory. Our main purpose is to study how this mechanism affects the evolution of cooperation, and if it really facilitates cooperation, what supports the promotion phenomenon. By means of systematic Monte Carlo simulations, we show that, actually as what we expect, this mechanism can promote the evolution of cooperation. Interestingly, we find that this promotion phenomenon will be supported by a recovery effect, which can be well interpreted by an evolution resonance of standard deviation of fitness coefficient. Moreover, we examine the impact of different levels of uncertainty by strategy adoptions. In the remainder of this paper we will first describe the considered evolutionary game, subsequently we will present the main results, and finally we will summarize our conclusions.

\section{The Model}

We consider an evolutionary prisoner's dilemma game with players located on the sites ($x$) of a regular $L \times L$ square lattice with periodic boundary conditions. Initially, the player on the site $x$ is designated as a cooperator ($s_x=C$) or defector ($s_x=D$) with equal probability, and is distributed a fitness coefficient $\phi_x(t)$ as well. As characterized by previous literatures \cite{33,35}, we use the rescaled payoff matrix: the temptation to defect $T=b$ (the highest payoff received by a defector if playing against a cooperator), reward for mutual cooperation $R=1$, and both punishment for mutual defection P and the sucker's payoff S (the lowest payoff received by a cooperator if playing against a defector) equaling to 0, whereby $1 \le b \le 2$ ensures a proper payoff ranking. The game is iterated in accordance with the Monte Carlo (MC) simulation procedure comprising the following elementary steps. First, a randomly selected player $x$ acquires its payoff $P_x$ by playing the game with its four nearest neighbors, and evaluates its fitness $\Pi_x$ by following expression
\begin{equation}
 \Pi_x=\phi_x(t) \times P_x.
\end{equation}
Next, one randomly chosen neighbor $y$ also obtains its fitness $\Pi_y$ by the same way as player $x$. Lastly, player $x$ adopts the strategy from the randomly selected player $y$ in accordance with the probability,
\begin{equation}
W(s_y \to s_x)=\frac{1}{1+exp[(\Pi_x-\Pi_y)/K]},
\end{equation}
where $K$ denotes the amplitude of noise or its inverse ($1/K$), the so-called intensity of selection \cite{19}. In one full Monte Carlo step (MCS), each player $x$ has a chance once on average to adopt a strategy from the randomly selected neighbor as described above. Interestingly, the fitness coefficient $\phi_x(t)$ is also updating according to the following protocol. Initially, each player $x$ possesses the same fitness coefficient $\phi_x(t)=1.0$ to avoid preferential influence before the game. Then, if player $x$ adopts the strategy of neighbor $y$, which is similar to a failure among the competitions or battles, its fitness coefficient will decrease according to $\phi_x(t+1)=(1- \alpha)* \phi_x(t)$, where $\alpha$ ($\alpha \ll 1.0$) is the punishment rate. Otherwise, the fitness coefficient will keep constant. Importantly, we assume that the step when player $x$ is punished is regarded as the first step. If player $x$ keeps its strategy changeless in the subsequent $M$ steps (called memory length), which is like remaining indefectible in the succedent competitions or battles, its fitness coefficient will recover 1.0. However, once player $x$ still adopts strategy from others during one of the subsequent $M$ steps, that step will spontaneously be regarded as the first step over again. The above process will be iterated till its fitness coefficient gets back to 1.0. It is notable that these setups can be interpreted from the social viewpoints, defected individuals among the competitions or battles lose part of their own properties or territories, but importantly they try to redeem their loss through a certain period of struggle.

Results of Monte Carlo simulations presented below were obtained on populations comprising $100 \times 100$ to $400 \times 400$ individuals, whereby the fraction of cooperators $\rho_C$ was determined within last $10^4$ full steps of overall $2 \times 10^5$ MCS. In order to overcome the influence of large $M$ value, longer transient time was discarded. Moreover, since the recorded punishment may introduce heterogeneous influence of fitness coefficient, final results were averaged over 20 to 40 independent runs for each set of parameter values in order to assure suitable accuracy.

\section{Simulation Results and Discussion}

We start by examining the effect of the above introduced punishment rate $\alpha$ on the evolution of cooperation while keeping a certain memory length $M$ constant. Figure 1 shows how $\rho_C$ varies in dependence on the temptation of defect $b$ for different values of $\alpha$. Evidently, $\alpha=0$ (irrespective of the value of $M$) recovers the traditional version of spatial prisoner's dilemma game, where cooperators will die out even if the value of $b$ is small. However, as $\alpha$ increases, the evolution of cooperation will be promoted more effectively. It can be clearly observed that for $\alpha=0.02$ cooperation can be facilitated to a near-complete dominant strategy when the value of $b$ is small. With the continuous increment of $\alpha$, defectors can only exist when the temptation to defect is sufficient large. These results suggest that when the punishment is taken into account, the evolution of cooperation will thrive. While an increase of punishment rate $\alpha$ will directly result in an increase of cooperation, since the introduction of punishment may generate a heterogeneous state for the whole population during the process of evolution.

In what follows, we will explore the ability of memory length $M$ to facilitate and maintain cooperation. Results presented in figure 2 clearly shows $\rho_C$ in dependence on the whole relevant span of $b$ for different values of $M$. For $M=0$, it corresponds to the case where only punishment exists but no memory is contained in the game. Compared with the results of traditional version in figure 1, it is evident that better facilitation of cooperation could be warranted, which implies that the performance of merely spatial reciprocity has been improved \cite{31}. Interestingly, as the value of memory length $M$ increases, the evolution of cooperation will fare better, namely, the survivability of cooperators will monotonously enhance with increasing $M$. In particular, when the value of $M$ is sufficiently large, the promotion effect on cooperation will become nearly changeless with the increment of $M$, that is, the fraction of cooperators will converge to a deterministic value for enough large value of $M$. It is worth emphasizing that in such condition, the evolution of system will take longer time to arrive at the steady state. Moreover, it is also notable that compared with the case of merely punishment ($M=0$), cooperation could be better promoted within the framework involved with the joint influence of punishment and memory.

To further certify the above observations regarding the promotion of cooperation, we visually inspect the characteristic spatial patterns of cooperators and defectors for different situations. As illustrated in figure 3, in the original patterns ($t=0$) cooperators and defectors are evenly distributed on the lattice. For the traditional version which does not include punishment and memory (upper panel), cooperators will decrease soon and form sliced cooperator clusters. With the evolution of system, the few remaining cooperator clusters can not resist against the invasion of defectors, and defection will finally become the dominance strategy, which implies that only the network reciprocity among cooperators can not sufficiently work if the value of $b$ is relatively high \cite{31}. In other cases, when the punishment was introduced into the game, evident change can be observed (middle panel). We can see that cooperators will first decrease and form small clusters, then these cooperator clusters can insure appropriate environment for cooperation thriving through keeping their dynamic sizes nearly constant. Interestingly, based on the above achievements of punishment, the joint effect of punishment and memory on the evolution of cooperation is more distinct as lower panel of figure 3 illustrates. The few remaining cooperator clusters not only resist against the fast invasion of defection, but importantly, they start to recover the lost ground and take up the whole system. Consequently, the recorded punishment involved with punishment and memory can result in a recovery effect, which halts and eventually reverts the fast vanishing of cooperators toward their undisputed dominance.

It remains of interest to explain why cooperative behavior is promoted through a recovery effect. In order to provide answers, we investigate in figure 4 time courses for standard deviation of fitness coefficient $S(\phi)$ and fraction of cooperators $\rho_C$ under different situations that have been discussed above. As is well known, standard deviation denotes the deviation degree between individual values and average value of system. The larger the value of standard deviation, the more remarkable the heterogeneity among players. For the traditional version, the standard deviation will always equal zero (note that values of $S(\phi)$ were recorded in-between full Monte Carlo steps), which means that fitness coefficient is the same for each player. In such case, cooperators will be decimated and defection becomes the complete dominance strategy (note that values of $\rho_C$ were also recorded in-between full Monte Carlo steps). This is actually what we would expect, given that defectors are, as individuals, more successful than cooperators and will thus be chosen more likely as the potential strategy if $b$ is large. Interestingly, however, when the punishment was introduced, the above tide will change. As can be observed from the top panel of figure 4, in the most early stages of evolution process the standard deviation will first exhibit a weak peak and then keep a certain value nearly changeless over time. This implies that the heterogeneity among players has been formed within the system. Correspondingly, this weak peak will halt the decimation of cooperators in the early stages and turn to the fast spreading of cooperation, which also attests to the fact that heterogeneity plays an important role in the substantial promotion of cooperation \cite{29,32}. Quite surprisingly, when recorded punishment involved with memory and punishment was considered, we can observe that there exists an evident peak of standard deviation analogously to the so-called coherence resonance \cite{53}. While such an evolution resonance of standard deviation will effectively change the initial downfall of cooperators, and result in the faster widespread cooperation. The complete dominance of cooperation in turn accelerates the system returning homogeneous state, namely, the standard deviation will get back to zero again. Hence, we argue that the evolution resonance of standard deviation stimulates a recovery effect, which could promote the evolution of cooperation better than the existence of heterogeneity among players alone.

Finally, it is instructive to examine the evolution of cooperation under different levels of uncertainty by strategy adoptions. The latter can be tuned via $K$, which acts as a temperature parameter in the employed Fermi strategy adoption function \cite{19}. Accordingly, in the limit $K \to \infty$, all information is lost, switching to neighbor's strategy is like tossing a coin. While in the limit $K \to 0$, the strategy of selected neighbor is always adopted provided that its fitness is higher. Results of phase separation lines on the $K-b$ parameter plane are presented in figure 5, whereby below the lines mixed cooperators and defectors coexist, while above, a homogeneous defector state prevails. Notably, the phase transition of tradition version (black line) exists an optimal level of uncertainty for the evolution of cooperation, at which cooperators are able to survive at the highest value of $b$, as was reported in previous literatures \cite{27,53}. While this phenomenon can only be observed on interaction topologies lacking overlapping triangles \cite{54}. Interestingly, when the punishment (red line) and recorded punishment (blue line) were introduced, the qualitatively analogical phase transitions can be acquired. They exhibit the optimal values of $K$ as well, but obviously, the coexistence space of cooperators and defectors is substantial extended, which further support the above result that cooperation could be greatly facilitated under the joint effect of memory and punishment. Moreover, these qualitatively similar phase transitions imply that consideration of recorded punishment does not alter the initial interaction network, which is slightly different from some previous literatures \cite{28,54}. Since the square lattice obviously lacks overlapping triangles and enables the observation of an optimal of $K$.

\section{Conclusion}
In sum, we have studied the effect of recorded punishment involved with memory and punishment on the evolution of cooperation in the spatial prisoner's dilemma game. We show that recorded punishment is an effective mechanism to promote cooperation. With monotonously increasing punishment rate or memory length, cooperative behavior will be better promoted. Interestingly, if only punishment was introduced, cooperators could avoid the destiny of dying out. In the very early stages of evolution process, cooperators will first be decimated, then the few remaining cooperators would form small clusters to resist the invasion of defectors. With the evolution of game these clusters will become large and impervious to defector attacks even at high temptations to defect. Meanwhile, we show that punishment could result in the heterogeneous distribution of individual fitness coefficient, whereby the heterogeneity plays an important role in the promotion of cooperation \cite{29,32}. Further interesting is the fact that the consideration of recoded punishment could make cooperation thrive better, even to the complete dominance. While this facilitation phenomenon of cooperation has been attributed to a recovery effect, namely, in the very early stages of the game defectors are able to plunder very effectively, but the few remaining cooperators form clusters and recover the lost ground towards their undisputed dominance soon. Quite correspondingly, this recovery effect can be explained by an evolution resonance of standard deviation. When the standard deviation of individual fitness coefficient reaches a peak, which means an evident heterogeneous state within system, the initial downfall of cooperators will be effectively halted and turn to the fast spreading. When cooperation becomes the dominance strategy, the standard deviation of fitness coefficient will return zero which makes the system regain the homogeneous state. In addition, by exploring the phase transition lines, we further support the result that recorded punishment promotes cooperation better.

Since recorded punishment seems very reasonable and very widely applicable as well realistically justifiable. We hope that it can inspire further studies, especially provide the theoretical instruction to some social dilemmas.

\section*{Acknowledgements}

Zhen Wang acknowledges support from the Center for Asia Studies of Nankai University (Grant No. 2010-5) and from the National Natural Science Foundation of China (Grant No. 10672081).





\begin{figure}[htbp]
\begin{center}
\scalebox{0.35}[0.35]{\includegraphics{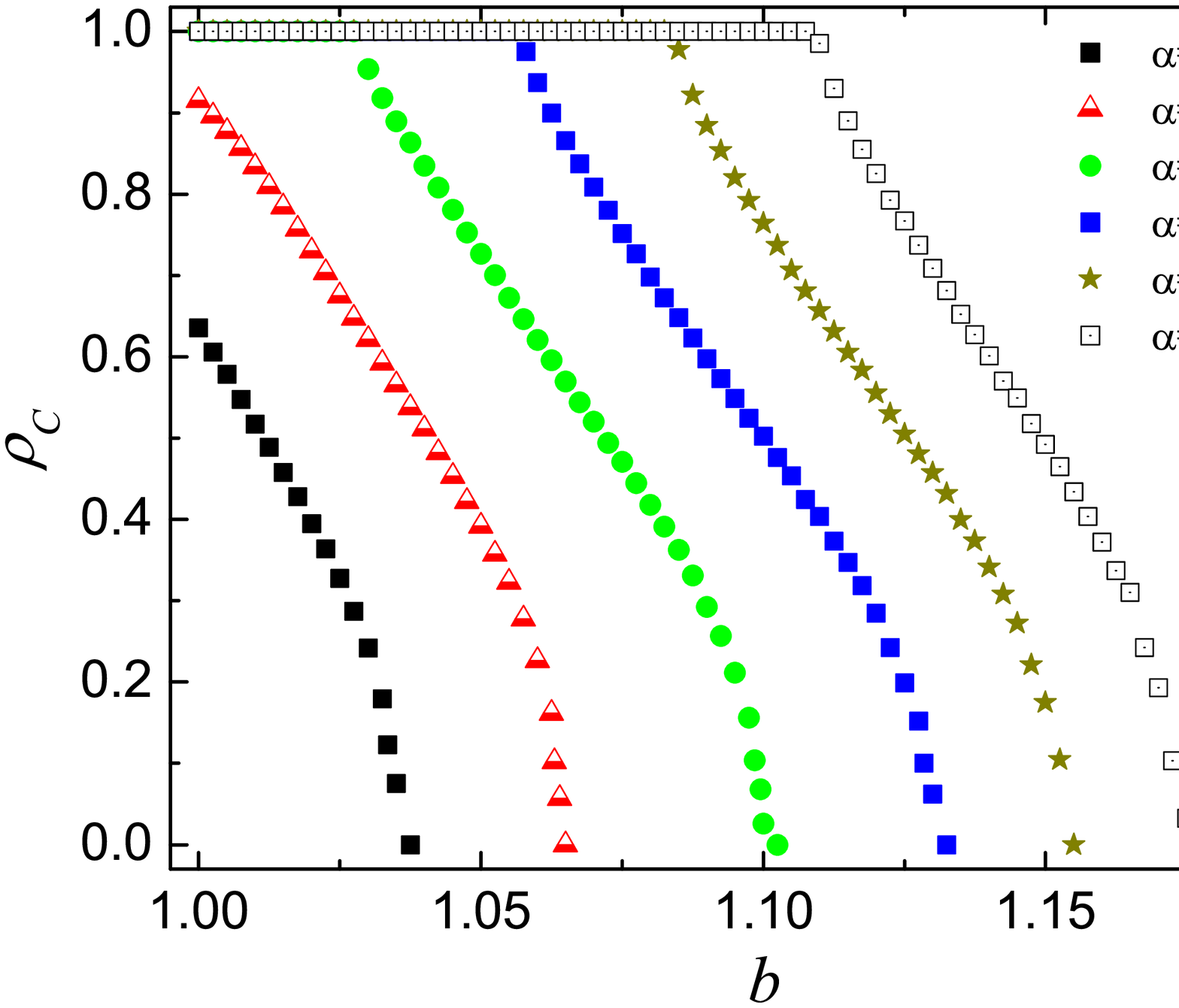}}
\caption{(\textit{color online}) Frequency of cooperator $\rho_C$ in dependence on the parameter $b$ for different values of punishment rate $\alpha$. Note that $\alpha=0$ will recover the traditional version, while the increment of $\alpha$ will introduce punishment into game. Depicted results were obtained for
$M=5$ and $K=0.1$.}\end{center}
\end{figure}

\begin{figure}[htbp]
\begin{center}
\scalebox{0.35}[0.35]{\includegraphics{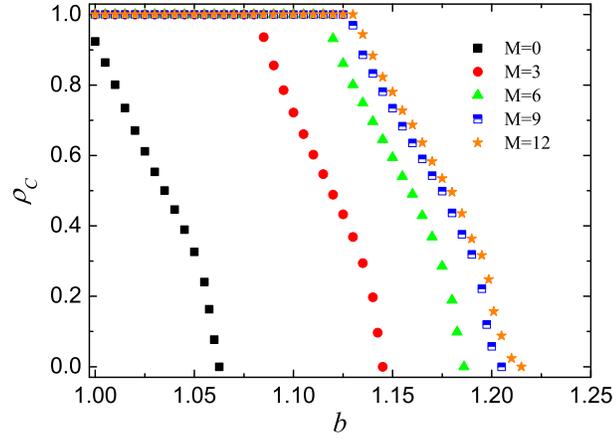}}
\caption{(\textit{color online}) Frequency of cooperator $\rho_C$ in dependence on the parameter $b$ for
different values of memory length $M$. Note that with increment of memory length $M$, cooperation
will be greatly facilitated. However, the enhancement of cooperation will reach a limit if memory
length is sufficient large. Depicted results were obtained for $\alpha=0.1$ and $K=0.1$.}
\end{center}
\end{figure}

\begin{figure}[htbp]
\begin{center}
\scalebox{0.5}[0.5]{\includegraphics{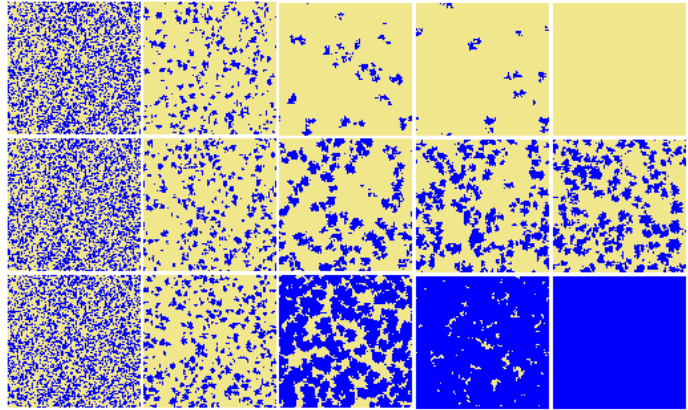}}
\caption{(\textit{color online}) Characteristic spatial patterns of cooperators (blue) and defectors (yellow) under different situations. Initially cooperators and defectors are distributed evenly, but they evolve into different patterns. {\it Top panel}: it does not include punishment and memory ($\alpha=0$ and $M=0$), {\it i.e.}, traditional version; {\it middle panel}: only consider the influence of punishment ($\alpha=0.1$ and $M=0$); {\it bottom panel}: both punishment and memory exist ($\alpha=0.1$ and $M=5$). Snapshots were given after $t=$0, 10, 100, 200, 10000 steps for all the panels. Depicted results were obtained for $b=1.05$ and $K=0.1$.}\label{1}
\end{center}
\end{figure}

\begin{figure}[htbp]
\begin{center}
\scalebox{0.45}[0.45]{\includegraphics{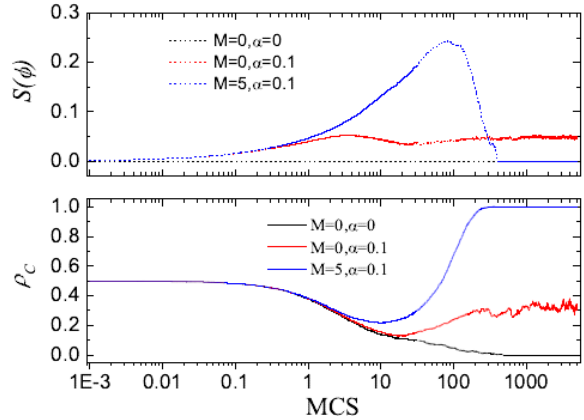}} \caption{(\textit{color online})
Time courses depicting standard deviation of individual fitness coefficient
$S(\phi)$ and fraction of cooperator $\rho_C$ for three different situations. {\it Top panel}: it depicts
the evolution of standard deviation, which reflects the distribution of payoff coefficient $\phi_x(t)$
among players. Note that when recorded punishment ({\it i.e.}, $\alpha=0.1$ and $M=5$) is considered, there exists an evolution resonance of standard deviation. {\it Bottom panel}: it characterizes the evolution of
cooperation. With respect to the recorded punishment, the evolution resonance of standard deviation will
lead to an obvious reverse of cooperation to the dominance. Note that the horizontal axis is logarithmic
and that values of $S(G)$ and $\rho_c$ are recorded also in-between full Monte Carlo steps (MCS) to ensure
accuracy. Depicted results were obtained for $b=1.05$ and $K=0.1$. and $K=0.1$ on a $400 \times 400$ square lattice.
}
\end{center}
\end{figure}

\begin{figure}[htbp]
\begin{center}
\scalebox{0.35}[0.35]{\includegraphics{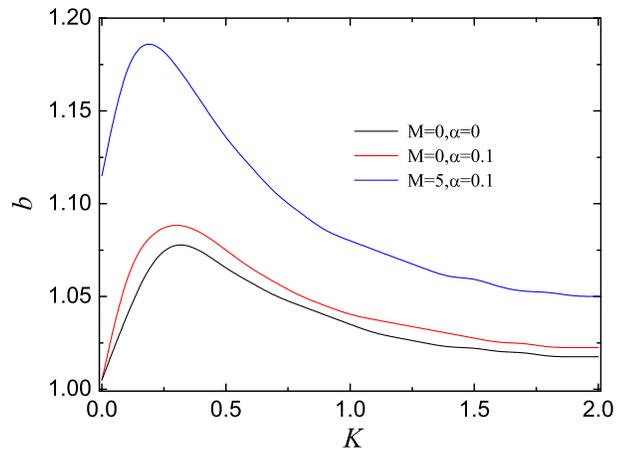}}
\caption{(\textit{color online})Phase separation lines on the K-b parameter plane for different situations
on the square lattice. Lines denote the border separating mixed $C+D$ (below) and pure defector $D$
states. Note, irrespective of which kind of situations, there exists an intermediate uncertainty in
the strategy adoption process (an intermediate value of $K$) for which the survivability of cooperators
is optimal. But when the recorded punishment ({\it i.e.}, $\alpha=0.1$ and $M=5$) is considered, the existence of optimal uncertainty for cooperation is most obvious. D $\leftrightarrow$ C+D transition is qualitatively different.}
\end{center}
\end{figure}

\end{document}